\documentclass[12pt]{article}
\pdfoutput=1

\usepackage{copyrightbox}

\usepackage{amsfonts,amssymb,color,cancel,mathtools,setspace,tabularx,subfigure}
\usepackage[,textheight=8.6in,textwidth=7in,letterpaper]{geometry}

\usepackage{suffix}
\usepackage{mathtools}
\DeclarePairedDelimiterX\MeijerM[3]{\lparen}{\rparen}%
{\begin{smallmatrix}#1 \\ #2\end{smallmatrix}\delimsize\vert\,#3}

\newcommand\MeijerG[8][]{%
  G^{\,#2,#3}_{#4,#5}\MeijerM[#1]{#6}{#7}{#8}}

\WithSuffix\newcommand\MeijerG*[7]{%
  G^{\,#1,#2}_{#3,#4}\MeijerM*{#5}{#6}{#7}}

\usepackage{hyperref}
\hypersetup{
 colorlinks=true, 
    linktoc=page,     
    linkcolor=blue,
}

\begin{document}


\title{{\bf  \LARGE Signatures of Extra Dimensions\\ in Black Hole Jets}}

 \author{Alexandra Chanson$^{1}$, {\large Maria J. Rodriguez}$^{1,2}$,  \\
 $^{1}${\small Department of Physics, Utah State University, 4415 Old Main Hill Road, UT 84322, USA.}\\
 \\ 
  $^{2}${\small  Instituto de Fisica Teorica UAM/CSIC, Universidad Autonoma de Madrid,} \\{ \small13-15 Calle Nicolas Cabrera, 28049 Madrid, Spain.}\\
}
%
%

\maketitle

\vspace{2cm}

\abstract
\vspace{1cm}
One of the leading mechanisms powering relativistic black hole jets is the Blandford–Znajek (BZ) process. Inspired by its success we construct energy extracting models for black holes in five space-time dimensions. Here, we find solutions to the force-free electrodynamic equations representing plasma-magnetospheres for slowly rotating Myers-Perry black holes. Both, energy and angular momentum fluxes are computed for these solutions realizing power extraction from black holes in higher dimensions. Comparisons of the main features of the five-dimensional BZ models with lower four-dimensional counterparts are discussed.

 \vspace{4cm}
\footnotetext{$^\dagger$a.chanson@usu.edu}
\footnotetext{$^\heartsuit$majo.rodriguez.b@gmail.com}

\vspace{2cm}


 \tableofcontents

\section{Introduction}
\label{sec:Introduction}

Several types of powerful high-energy events illuminate the universe. Most have been observationally associated with central black holes of some active galaxies, quasars, and also by galactic stellar black holes, neutron stars or pulsars. For black holes, it is widely believed that these are spinning and threaded by large scale astrophysical magnetic fields which helps convert binding and rotational energy in a highly efficient outflow process. Our understanding of these systems benefits hugely from progress on the observation front \cite{Kovalev:2019cue,Asada_2012,EventHorizonTelescope:2021iqj}, the sophisticated general relativistic magnetohydrodynamics simulation \cite{Komissarov:2008ic,Davis:2020wea} and new insights from theoretical efforts \cite{Lupsasca:2014pfa,Camilloni:2022kmx,Armas:2018atq,Jacobson:2017xam} that seek to unravel these fascinating phenomena.

In the 70's Blandford and Znajek layed out a successful model for energy and angular momentum extraction from rotating Kerr black holes surrounded by magnetospheres  \cite{Blandford:1977ds}. These magnetospheres are described by the highly nonlinear equations of force-free electrodynamics, or FFE.  Over the past three decades, the BZ process has been extensively studied in General Relativity (GR). One of the basic predictions of the BZ model is that for slowly rotating black holes the total electromagnetic energy flux extracted from the black hole scales as the spin $\alpha$ squared
\begin{eqnarray}
P\sim \alpha^2
\end{eqnarray}
Numerical \cite{McKinney_2005,Hawley:2005xs} and analytical work (see e.g. \cite{Camilloni2022} and refrecens therein) hints that a much steeper dependence of power on $\alpha$ occurs as the black hole spin increases. While it is clear that the  power from the black holes does depend on the spin parameter, other effects such as collimation introduce an even stepper behavior on $\alpha$ in the power spectrum.

The astrophysical problem is therefore slightly different than the plain BZ model, since what we actually observe are jets and not an isotropic outflow. A collimating agent is required for the black hole jet models. For the collimated morphology, the agent translates into specific boundary conditions instead of the bare black hole geometry. Examples include paraboloidal boundaries along the wind from the surrounding accretion disks \cite{Tchekhovskoy_2010}, or in the form of a background vertical magnetic field held in place by a disk far from the black hole \cite{Palenzuela:2010nf,Palenzuela:2010xn,Moesta:2011bn}.\\
The proliferation of sensitive telescopes such Event Horizon Telescope, are providing a wealth of information to reveal the launching and initial collimation region of extragalactic radio jets in for example Messier 87 \cite{Blandford:2022ewb} and Centaurus A \cite{EventHorizonTelescope:2021iqj}-- the closest radio-loud source to Earth. This ever-increasing number of observations so far have shown to be consistent with the predictions from GR. In the era to come there will be growing interest in attempting to show, or to exclude, the possibility that the results of observations may be better described by some alternative theory of gravity. As we divert from GR we need to consider alternative theories, including higher dimensions \cite{ Cardoso:2019vof,Banerjee:2019nnj,Vagnozzi:2019apd,Sperhake:2013qa}.

Here, we address the following main questions: what is the effect in black hole jet models of the number of spacetime dimensions? In particular, can a black hole in five space-time dimensions (5D) support energy extracting magnetospheres? Can higher dimensions introduce in the jet power an even steeper scaling on the spin parameter?

In this context, we construct energy extracting models for black holes jets in 5D spacetimes. These are solutions of the force-free electrodynamic equations in a slowly rotating $5D$ Myers-Perry black hole \cite{Myers:1986un}. Even-though they do not realize in nature, they are worth exploring, since they have interesting features in the vast landscape of alternative theories of GR. One of the remarkable properties of higher-dimensional gravity is its connection with other areas of physics. While the physics of higher-dimensional BZ models that we develop is a matter of interest per se, these may play a central role in holography, the microscopic interpretation of the Bekenstein–Hawking entropy \cite{Strominger:1996sh}, and direct contact with experiments such as heavy-ion collisions performed at the LHC \cite{LHeC:2020van} and EIC \cite{Accardi:2012qut}. Moreover, upon dimensional reduction, such black hole jet models can also be relevant in astrophysics and also understanding how special our $4D$ world is in the space of all possible values.

This paper is organized as follows: basic equations governing stationary force-free fields around 5-dimensional Myer-Perry black holes with one spin are introduced, and for the first time we set-up the BZ model in higher dimensions in Section 2. We present new exact solutions of the FFE equation in 5D flat spacetimes and static black holes in Section 3. In Section 4 we describe a perturbative approach to obtain a self-consistent highly collimated jet solutions. Physical properties, such as energy extraction rate are also discussed and compared with the 4D solution. Discussions are given in Section 5.

\section{FFE for $5D$-black holes}
\label{Sec:MyersPerry}

In this section we review conventions and equations for an extension of the BZ model by promoting the FFE equations to 5D space-times. We also collect some results on rotating Myer-Perry 5D black holes in order to make the paper more self-contained. \\

Force-free electrodynamics describes systems in which most of the energy resides in the electromagnetic (EM) sector of the theory where the energy-stress tensor $T_{\mu\nu}\sim T^{EM}_{\mu\nu}$. This approximation is known as the “force-free” condition. Here we argue that these are generalized straightforwardly to higher dimensional spacetimes. We hence want to find stationary and axisymmetric solutions of the following FFE equations
\begin{equation}\label{eq:FFE}
F_{\mu\nu}\,J^{\nu}=0\,,
\end{equation}
 supplemented by Maxwell's equations
\begin{eqnarray}
F_{\mu\nu}\,J^{\nu}&=0\,,\label{eq:maxwell1}\\
\nabla_{\nu}F^{\mu\nu}&=J^{\mu}\,.\label{eq:maxwell2}
\end{eqnarray}
where $\mu,\nu=\{t,\phi,\psi,r,\theta\}$. We motivate our 5D FFE model from observation that standard (four-dimensional) Einstein gravity can straightforwardly be generalized to higher dimensions. 5D FFE is similar in spirit to BZ \cite{Blandford:1977ds} but differs in the way the indices $\mu,\nu \in D$ and on the choice of black hole background. Higher dimensional gravity exhibits a much richer dynamics due to the existence of extended black objects with a same mass and angular momentum but with different horizon topologies such as black string, black rings and over-spinning spherical black holes \cite{Black zoo}. The set up for our computation will focus only on a 5D rotating black hole background solution, supported by an appropriate set of boundary conditions follows next.\\

We shall explicitly work in the single spinning 5D-Myers-Perry black hole \cite{Myers:1986un} solution for the background metric where certain calculations are
simplified. The Myers-Perry metric, parametrized by the mass $m$, and (one) rotation $a$, takes the form
\begin{eqnarray}
ds^2&=& -dt^2 + \frac{m}{\Sigma}\left( dt-a\sin^2\theta
\,d\phi\right)^2 +{\Sigma\over\Delta}dr^2+\Sigma d\theta^2
+(r^2+a^2)\sin^2\theta\, d\phi^2 \nonumber\\
&&
+ r^2\cos^2\theta\, d\psi^2\,,
\label{mphole}
\end{eqnarray}
where
\begin{eqnarray} \Sigma=r^2+a^2\cos^2\theta\,,\qquad
\Delta=r^2+a^2-m\,. 
\end{eqnarray}
The event horizon is $r\equiv r_H = \sqrt{m-a^2}$ and the angular velocities $\Omega^H_{\phi}= a/(r_H^2+a^2)$ and $\Omega^H_{\psi}=0$. Unlike Kerr black holes, the Myers-Perry black hole with a single spin is not bounded. In this black hole 5D space, the extremal value $m=a^2$ corresponds to an irregular solution where the horizon is destroyed, leading to naked singularity solution. In this Boyer-Lindquist type coordinates, this metric is square diagonal, and we can identify 
\begin{equation}
ds^2= (g_1)_{ab}\,dx^{a}dx^{b}+(g_2)_{\alpha\beta}\,dx^{\alpha}dx^{\beta}
\end{equation}
where $\{x^{a,b}=t,\phi,\psi\}$ and $\{x^{\alpha,\beta}= r,\theta\}$. 
In such backgrounds it can be easily shown that all stationary and axisymmetric  field strengths, solutions to $(\ref{eq:FFE})-(\ref{eq:maxwell2})$, are of the form
\footnote{Note that the most general stationary and axisymmetric gauge field is
\begin{equation}
A=A_{a}(x_1,x_2) \,dx^a+A_{\alpha}(x_1,x_2) \,dx^{\alpha} \,.
\end{equation}
hence defining the fluxes by $\Psi_{\phi}\equiv 1-A_{\phi}$ and $\Psi_{\psi}\equiv 1-A_{\psi}$, the electromagnetic field angular velocities $ \omega_{\phi,\psi}$ taking $-\partial_{x_{\alpha}} A_t =\omega_{\phi}\, \partial_{x_{\alpha}} A_{\phi} +\omega_{\psi}\, \partial_{x_{\alpha}} A_{\psi} $ and current $I=\left(\sqrt{ \frac{\det g_2}{-\det g_1}}\right)^{-1}\left(\partial_{x_1}A_{x_2}-\partial_{x_2}A_{x_1}\right)$ one finds via $F\equiv dA$ the corresponding the general expression for the field strength (\ref{eq:fieldstrength}).}
\begin{equation}\label{eq:fieldstrength}
F=d\Psi_{\phi}\wedge [d\phi-\omega_{\phi}\,dt]+d\Psi_{\psi}\wedge [d\psi-\omega_{\psi}\,dt]+I\,\sqrt{ \frac{\det g_2}{-\det g_1}}\,dr\wedge d\theta\,.
\end{equation}
where the fluxes $\Psi_{\phi,\psi}$, the electromagnetic field angular velocities $\omega_{\phi,\psi}$ and current $I$ are functionals of an arbitrary function $f(r,\theta)$. We therefore argue that the BZ process in 5D has five free parameters. These functions encode the energy and angular momentum fluxes. For a stationary and axisymmetric FFE field $(\ref{eq:fieldstrength})$ the corresponding total energy flux can be defined by
\begin{equation}\label{power}
{P}\equiv \int  \, T^{r\nu}_{EM} (\xi_{t})_{\nu}\, dV=-2 \,(2\pi)^2\int (\omega_{\phi} \,\partial_{\theta}\Psi_\phi + \omega_{\psi} \,\partial_{\theta}\Psi_\psi) \, I \, d\theta\,,
\end{equation}
and the angular momentum fluxes
\begin{eqnarray}\label{angpower}
{L}_{\phi}\equiv -\int T^{r\nu}_{EM} (\xi_{\phi})_{\nu} \,dV=-2 \,(2\pi)^2\int I \,\partial_{\theta}\Psi_\phi  \, d\theta \,,\\
{L}_{\psi}\equiv -\int T^{r\nu}_{EM} (\xi_{\psi})_{\nu} \,dV=-2 \,(2\pi)^2\int I \,\partial_{\theta}\Psi_\psi  \, d\theta\,.
\end{eqnarray}
We work in the souther hemisphere $0\le\theta\le \pi/2$ and assume a reflection-symmetric magnetosphere along the equator. The factor of 2 accounts for the contributions from both hemispheres, and the $(2\pi)^2$ contribution arises form the integration along the azimuthal  $(\phi,\psi)$-directions.
Here $T_{EM}^{\mu\nu}=F^{\mu\alpha}F_{\alpha}^{\nu}-(1/6)\,g^{\mu\nu}F_{\alpha\beta}F^{\alpha\beta}$ is the electromagnetic energy momentum tensor, $dV$ is the total space-time volume form and the timeline and two axial Killing vectors are $\xi_{t}^{\nu}=(1,0,0,0,0)$, $\xi_{\phi}^{\nu}=(0,0,0,1,0)$ and $\xi_{\psi}^{\nu}=(0,0,0,0,1)$ respectively.  While there is no sharp distinction between the azimuthal directions $\phi$ and $\psi$, working explicitly in 5D black hole space (\ref{mphole}) can simplify solutions and integrals over the space-time. In this paper, for simplicity we will further reduce the problem were the EM variables on $\psi$-direction are turned off. This will guarantee that there is no angular momentum flux ${L}_{\psi}=0$.

Plugging the expression for the stationary electromagnetic field (\ref{eq:fieldstrength}) into (\ref{eq:FFE}) leads to a basic reduced system of two fundamental non-linear FFE equations. These, respectively, two fundamental equations are
\begin{equation}\label{eq:streamTWO}
F_{r\nu}\,J^{\nu}=0\,, \qquad F_{\theta \nu}\,J^{\nu}=0\,.
\end{equation}
Before diving into solving these equations, we now consider the location of two of the relevant black hole surfaces (the event horizon and ergosurface) where (\ref{eq:streamTWO}) may become irregular. This leads us to establish boundary conditions to guarantee regularity for FFE equations that coincide with the smoothness requirement of the Poynting vector $F^2$ on these same surfaces. The discussion on the other two surfaces in the problem, the so-called light surface is postponed to Section \ref{subsec:physicalProps}.

\subsection{Event horizon and ergoregion}
\label{sec:eventandergo}

Given the MP black hole we can define the event horizon radius for the background by equating the zeros of the equation $g^{rr}=0$. The resulting value of location of the event horizon yields $r\equiv r_H = \sqrt{m-a^2}$. For slowly rotating black holes, where the spin parameter is small $a<<1$
\begin{eqnarray}
\frac{r_{H}}{r_0}=1-\frac{1}{2}\alpha^2 +O(\alpha^3)\,.
\end{eqnarray}
where $r_0=\sqrt{m}$ and reduced spin parameter $\alpha=a/\sqrt{m}$.\\

The boundary of the ergoregion, or ergosurface, can simply be defined as the locus of spacetime points where the asymptotic timelike Killing field $\partial_t$ becomes null, i.e. where $g_{tt}=0$. In the Myers Perry black hole (\ref{mphole}) the ergospheres $r_{e}(\theta)$ are found at
\begin{eqnarray}
r_{e}=\pm\sqrt{m-a^2\cos^2\theta-b^2\sin^2\theta}\,.
\end{eqnarray}
In the the small spin regime $a<<1$ the radius of the horizon 
\begin{eqnarray}
\frac{r_{e}}{r_0}=1-\frac{1}{2}\alpha^2 \cos^2\theta+O(\alpha^3)\,.
\end{eqnarray}

Our analysis here focuses mainly on the event horizon and the ergosurface. While both surfaces play a central role in the BZ model, exploiting the analogy with pulsar magnetospheres, we will further consider the physical significance of the so-called Light Surfaces (LS; see below). These are surfaces where magnetic field lines as a geometric construct ‘rotate’ at the speed of light in the same or the opposite direction with respect to an observer co-rotating with the magnetosphere.
 
\subsection{Boundary conditions}

We now examine the boundary conditions for the 5D FFE model that we propose. Introducing the black hole background a set of functions $\Psi_{\phi,\psi}$, $\omega_{\phi,\psi}$ and $I$ satisfying equations $(\ref{eq:streamTWO})$ will characterize a consistent model of a force-free magnetosphere given a choice of appropriate boundary. Following the original work by Blandford and Znajek \cite{Blandford:1977ds} at the equator $(\theta=\pi/2)$ the fluxes $d\Psi_{\phi,\psi}$ are considered to be discontinuous and determined by surface currents in the disc. At the same time, on the surface of the black hole the so called Znajek smoothness conditions will apply. For the $5D$ singly rotating black hole the relevant conditions for the fluxes $\Psi_{\phi,\psi}$ on the black hole event horizon yield a relation between the field functions
\begin{equation}\label{eq:Znajek}
I^H(\Psi^H)=(\omega^H_{\phi}-\Omega^H_{\phi})\,\,\frac{(r_H^2+a^2)\,r_H \sin\theta \cos\theta}{r_H^2+a^2\cos^2\theta} \,\partial_{\theta}\Psi^H_{\phi}\,.
\end{equation}
where $-\omega^{H}_{\phi}\equiv {\partial_r A^H_t}/{\partial_r A^H_{\phi}}=\partial_{\theta} A^H_t /{\partial_{\theta} A^H_{\phi}}$. This smoothness condition can be derived either from requiring regularity of the FFE equations on the vent horizon or, equivalently, from regularity of $F^2$. Since we are working with the equations $(\ref{eq:streamTWO})$ in single spinning MP black holes backgrounds, we see no contribution at the event horizon $\Psi^H_{\psi}$ in the equations. Therefore, without loss of generality we pick $\omega^H_{\psi}=0$. Notably the smoothness condition obey a similar structure to its 4D analog in Kerr.\\

At infinity $r\rightarrow\infty$ setting boundary conditions is more subtle, but as argued in \cite{Blandford:1977ds}, we resort to match the fields to solutions of flat space. 
The regularity condition in the region far from the black hole is
\begin{equation}\label{eq:Znajek}
I^{\infty}(\Psi^{\infty})=\pm \, \omega^{\infty}_{\phi} r \sin\theta \cos\theta \,\partial_{\theta}\Psi^{\infty}_{\phi}\,.
\end{equation}
To our knowledge solutions of FFE equation in flat vacuum five space-time dimensions are not known. Hence we now establish some relevant equations and 
derive a tower of solutions for the FFE equations. These will include not only solutions of the FFE equations in $D=5$ flat spacetime backgrounds, but also static black hole backgrounds.

\section{FFE Solutions in Flat and Static $5D$ Backgrounds}

In this section we compute exact solutions to the FFE equations in $D=5$ flat spacetime and static black holes backgrounds.
We work with black hole backgrounds containing a spherical event horizon.
For static black holes metric, one can simply set the spin parameter to zero $a=0$ in $(\ref{mphole})$. This corresponds to $5D$ metric found by Tangherlini \cite{Tangherlini:1963bw}. The black hole resembles Schwarzschild's solution not only due to the spherical event horizon topology but also on realizing the unique regular static solution of five-dimensional GR in vacuum. Static black holes will exhibit no energy extraction, however the new exact solutions to vacuum Maxwell equations in a Tangherlini metric that we now construct will be key for the FFE solutions for rotating $5D$ black hole that we present in a later section.

\subsection{Flat Spacetimes Solutions}

As in four space-time dimensions for Schwarzschild metric, in five spacetime dimensions there is a similar 'no hair theorem' that we can formulate \cite{BeskinBook}.\\

{\it Theorem: Non-rotating higher dimensional black holes cannot have magnetic fields themselves; the electric field in the exterior of a non rotating black hole must coincide with the field at a point located in the center.}\\

It is not in the scope of this paper to prove this theorem, but we will simply assume its validity since by solving the Maxwell's equations $(\ref{eq:maxwell1})$ and $(\ref{eq:maxwell2})$ we now find for $(\ref{eq:fieldstrength})$ with $I=\omega_{\phi}=\omega_{\psi}=0$ that the equations reduce to
\begin{eqnarray}\label{eq:operators1}
\mathcal{L}_{\phi} \Psi_{\phi}&\equiv&r \,\partial_r\left[\left(\frac{r^2-m}{r}\right)\partial_r \Psi_{\phi} \right] +\tan\theta \,\partial_{\theta}\left[\frac{1}{\tan\theta}\,\partial_{\theta} \Psi_{\phi} \right] =0\,,\\
\mathcal{L}_{\psi} \Psi_{\psi}&\equiv& r \,\partial_r\left[\left(\frac{r^2-m}{r}\right)\partial_r \Psi_{\psi} \right] +\cot\theta \,\partial_{\theta}\left[\frac{1}{\cot\theta}\,\partial_{\theta} \Psi_{\psi} \right]=0\label{eq:operators2}
\end{eqnarray}
and the electric and magnetic equations are separated. 

Besides, one can regard the field is degenerate $F \wedge F=0$ when 
\begin{eqnarray}
\frac{\partial_r \Psi_{\phi} }{\partial_{\theta} \Psi_{\phi}}=\frac{\partial_r \Psi_{\phi} }{\partial_{\theta} \Psi_{\phi}}\,.
\end{eqnarray}
This condition, that in four spacetime dimensions is necessary to solve the FFE equations, in 5-dimensions it is not.

In turn, equations (\ref{eq:operators1}) and (\ref{eq:operators2}) imply that the black hole magnetospheres can occur only in the presence of external electric currents. We can seek for solutions to these equations assuming
\begin{eqnarray}\label{eq:monopole}
\Psi_{\phi}=1-\sum_{l=0}^{\infty} c_l\, R_{l}(r)\, T^{\phi}_{l}(\theta)\,,\qquad \Psi_{\psi}=1-\sum_{n=0}^{\infty}d_n \,R_{n}(r) \,T^{\psi}_{n}(\theta)\,.
\end{eqnarray}
and arbitrary constants $c_l, d_n$.
Note that these operators differ slightly from flat space -- with no black hole $ m=0$ -- by a factor $({r^2-m})/{r}$ in the radial derivatives \footnote{ The radial eigenfunctions in 5D space-time become $R^{flat}_{k}(r)= r^{\pm k}$. } and therefore $T^{\phi}_{l}(\theta)$ and $T^{\psi}_{n}(\theta)$ will remain unchanged. Let us consider the eigenfunctions to the problem next.

\subsubsection*{Eigenfunctions}

The operators $\mathcal{L}_{\phi}$ and $\mathcal{L}_{\psi}$ are second order therefore there is a set of two eigenfunctions that solve $(\ref{eq:operators1})$-$(\ref{eq:operators2})$. Via separation of $(r,\theta)$-variables one can find the explicit solutions, each with separation constants $k=l,n$.

\begin{figure*}
\centering
\subfigure[\ vertical]{
	\label{fig:vertical}
	\includegraphics[width=4cm]{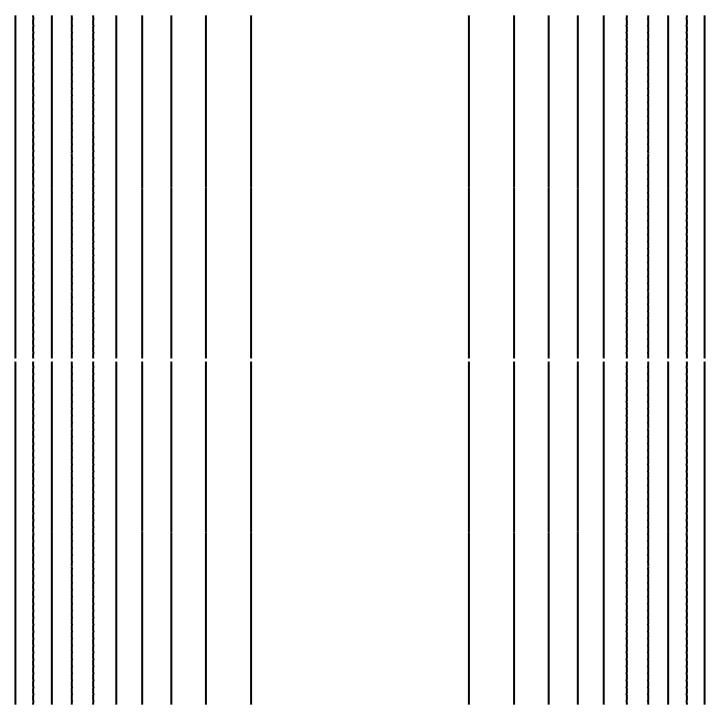}
}
\subfigure[\ radial]{
	\label{fig:radial}
	\includegraphics[width=4cm]{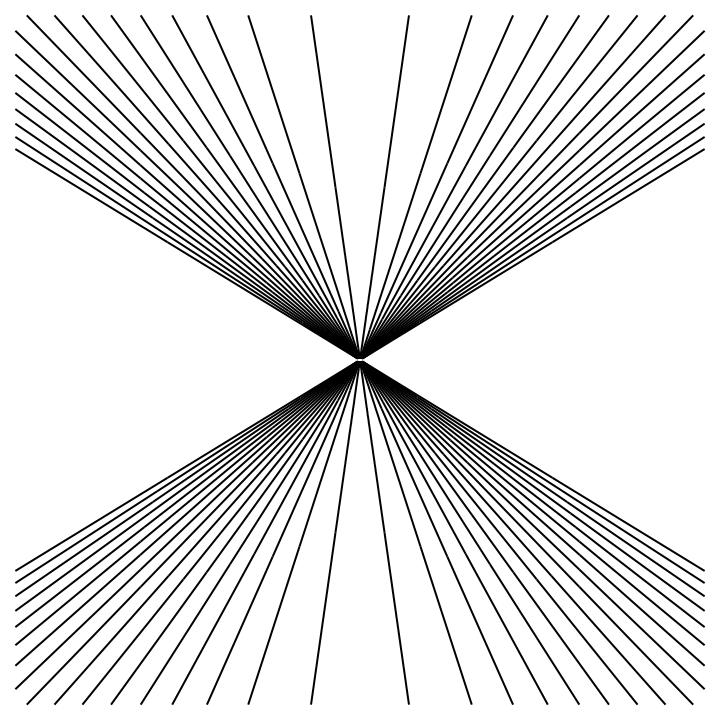}
}
\subfigure[\ parabolic]{
	\label{fig:parabolic}
	\includegraphics[width=4cm]{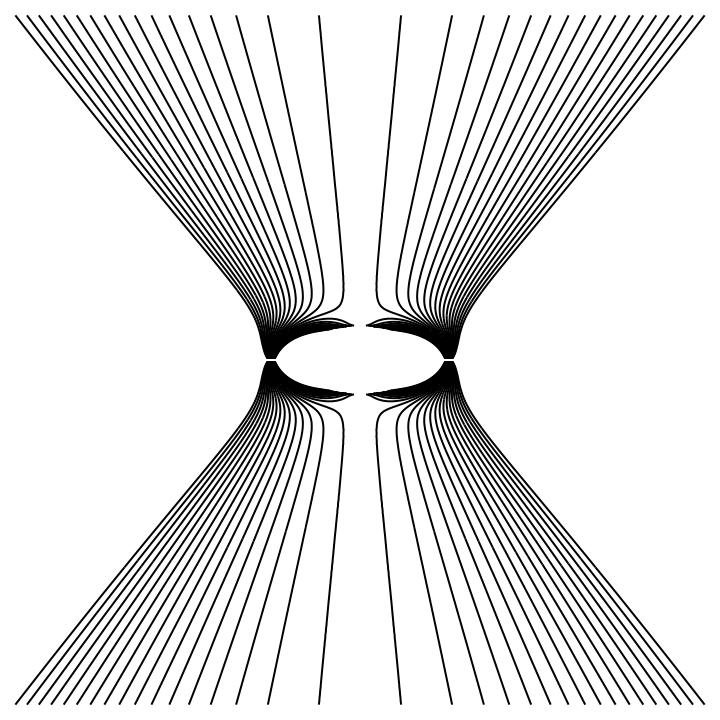}
}
\subfigure[\ hyperbolic]{
	\label{fig:vertical}
	\includegraphics[width=4cm]{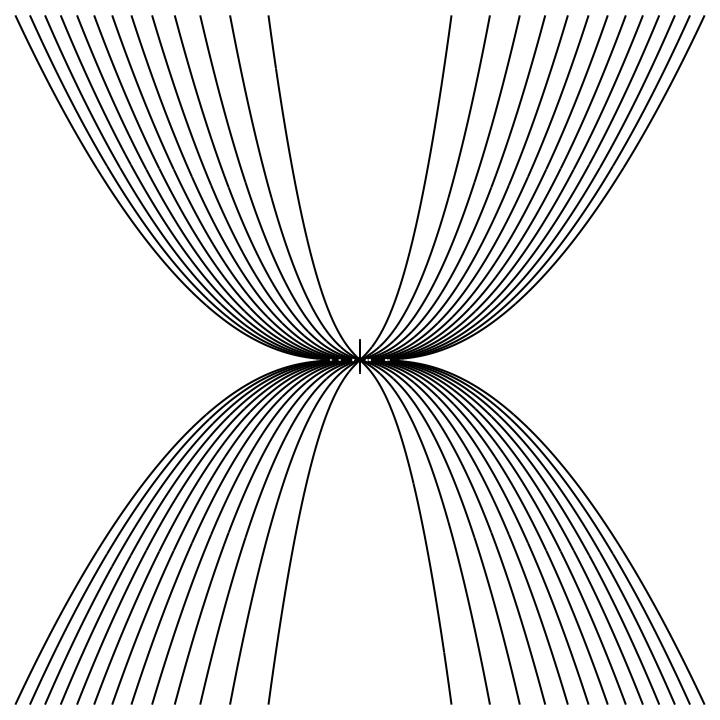}
}
\caption{Field lines of families of exact force-free magnetospheres in $5$-dimensional static black hole backgrounds. Our new solutions include the vertical, radial, parabolic, hyperbolic field geometries.}
\label{fig1}
\end{figure*}

 In more detail these are 
\begin{eqnarray}\label{eq:parabolic}
R_{k}(r)=F_{k}(r^2/m)\,,\qquad  \text{for} \, \, \, k=l,n>0
\end{eqnarray}
for the radial functions in $(\ref{eq:monopole})$ and for the spherical solutions 
\begin{eqnarray}\label{eq:parabolic}
T^{\phi}_{l}(\theta)=F_{l}(\sin^2\theta)\,,\qquad T^{\phi}_{n}(\theta)=F_{n}(\cos^2\theta)\,,\qquad \text{for} \, \, \, l,n>0
\end{eqnarray}

where
\begin{eqnarray}\label{eq:parabolic}
F_{k}(x)=\begin{cases}
		x \, {_2F_1}\!\left[{1-\frac{k}{2}, \, \,1+\frac{k}{2};2;x}\right],\\
		\MeijerG*{2}{0}{2}{2}{1-k/2, 1+k/2}{0, \, \,1}{x}\,
	\end{cases}
\end{eqnarray}
Each value of the separation constants gives different profiles for the flux functions. For instance, the first three harmonics ($l=0,1,2$) generate linearly independent flux solutions
\begin{eqnarray}\label{eq:parabolic}
\Psi_{\phi}=\begin{cases}1-c_0\,\log(\cos^2\theta)\\
1-c_0\,\log(r^2-m)\log(\cos^2\theta)\\
1- c_0 \, EllipticE[1-r^2/m] (EllipticE[\sin^2\theta]-EllipticK[\sin^2\theta])\\
1- c_0 \, r^2 \sin^2\theta
\end{cases}\,,\qquad 
\end{eqnarray}
Here we have only included the open field line solutions. In fact, the first two cases above correspond to the $l=0$ harmonic value. One can represent in cylindrical coordinates \footnote{Here the cylindrical coordinates we employed to represent the flux lines are $r=\sqrt{\rho^2+z^2}$ and $\tan\theta=\rho/z$.} the constant flux lines in a 2-dimensional plane (see Fig.\ref{fig1}). Similar profiles can be found for $\Psi_{\psi}$.

\subsection{Static 5D Black Holes Solutions}

An exact solution to force-free equations in the static Tangherlini $5D$ black hole space-time is 
\begin{eqnarray}
\Psi_{\phi}&=&1-c_1 \log(\cos\theta)\,,\qquad \omega_{\phi}=c_2 (\sin\theta)^{-2}\,, \qquad I_{\phi}=c_3\,, \\
\Psi_{\psi}&=&0\,,\qquad\qquad\qquad\qquad \omega_{\psi}=0\,, \qquad\qquad\qquad  I_{\psi}=0\,.
\end{eqnarray}
where $c_i$, $i=1,2,3$ are constants. 
 In all, these functions lead to the field strength
\begin{eqnarray}
F= \,d\theta \wedge\left[ c_1 \tan\theta \, d\phi-\frac{1}{\sin\theta\cos\theta}\,\left(\frac{c_3 \, r}{r^2-m}dr\pm  c_1 c_2 \, dt \right)\right]\,.
\end{eqnarray}
The expression also includes the exact solution in 5D flat space-time for $m\rightarrow 0$.

Further imposing Znajek's smoothness condition (\ref{eq:Znajek}) on the horizon constrains 
\begin{eqnarray}
c_3= - c_1 c_2 \sqrt{m}
\end{eqnarray}
 In all, these functions lead to the field strength
\begin{eqnarray}
F=c_1 \, d\theta \wedge\left[ \tan\theta \, d\phi-\frac{c_2}{\sin\theta\cos\theta}\,\left(\frac{r\sqrt{m}}{r^2-m}dr\pm dt \right)\right]\,.
\end{eqnarray}
In this section we found all solutions to vacuum Maxwell equations for static black holes in five space-time dimensions. In the following section we present new solutions to the FFE equations in $5D$ for slowly rotating black holes employing a perturbation technique. 

\section{FFE configurations for rotating 5D Black Holes}
\label{Sec:SlowRotation}

In this section we find for the first time solutions of FFE for rotation Myers-Perry black hole. To solve these equations  we resort to perturbation theory. More concretely, we construct energy extracting solutions for slowly rotating black holes perturbative in powers of the reduced spin parameter $\alpha$. The perturbation method that we will implement was first developed in four dimensions by Blandford-Znajek \cite{Blandford:1977ds}. The main purpose of this section is to build upon the work done in \cite{BeskinBook,Pan:2014bja,Gralla:2015uta,Gralla:2015vta}. Starting with the solutions for the static black hole that we found in the previous section we will now consider perturbations in the small-spin regime $\alpha<<1$ for the slowly single rotating Myer-Perry black holes.

\subsection{Vertical Field}

Among the exact solutions we derived in the previous section, in the non-rotating 5-dimensional black hole background there exists a collimated, uniform, {\it vertical} magnetic field solution. 
\begin{align}\label{eq:homogenous}
\Psi_{\phi}^{(0)}(r,\theta)=  r^2 \sin^2\theta\,.
\end{align}
Here and below, we work in the northern hemisphere $0\le\theta\le\pi/2$ with the southern fields determined by reflection. The field line is of a highly collimated cylindrical shape and the geometry dependence of this solutions is displayed in Fig. 1. If the black hole is spinning, toroidal magnetic fields will be generated, and the magnetic cylinder will become twisted into a helically twisted structure. The FFE equations (\ref{eq:FFE}-\ref{eq:maxwell2}) are highly nonlinear, hence our strategy to find a collimated cylindrical shape twisted field solutions is to perturb the non-rotating black holes (\ref{eq:homogenous}) by allowing small spin black hole perturbation. To leading orders in rotational parameter $\alpha$ namely, the corresponding solution can be expressed,
\begin{align}\label{eq:perturbations}
\Psi_{\phi}&=\Psi_{\phi}^{(0)}(r,\theta)+\alpha^2\, \Psi_{\phi}^{(2)}(r,\theta)+O(\alpha^3)\,,\\
r_0\,\omega_{\phi}&=\alpha \,\omega^{(1)}(r,\theta)+O(\alpha^2)\,,\\
 I&=\alpha \,I^{(1)}(r,\theta)+O(\alpha^2)\,.
\end{align}
where the function $\omega^{(1)}(r,\theta)$ and $I^{(1)}(r,\theta)$ are both functions of the gauge field $\Psi_{\phi}^{(0)}(r,\theta)$ and should be of the form
\begin{align}
\omega^{(1)}=\omega^{(1)}(\Psi_{\phi}^{(0)}) \,\qquad I^{(1)}=I^{(1)}(\Psi_{\phi}^{(0)})\,.
\end{align}
Note that other different zeroth-order solutions, i.e., the radial, parabolic and hyperbolic field solutions could also be considered in 5D space-times. As a first study of the BZ energy extraction models in higher dimensions our analysis in this section will focus only on the vertical filed line configurations.

According to the field regularity equation on the black hole horizon (\ref{eq:Znajek}) for the current $I$ to be well-behaved on the horizon $r=r_H=r_0$ then 
\begin{align}
I^{(1)}(r_0^2 \sin^2\theta)=  2 \, r_0^2 \sin^2\theta \cos^2\theta  \, (1-\omega^{(1)})\,.
\end{align}
This equation can also be written in a simpler form as $r_0^2\,I^{(1)}(x) = {2 \,x \left(r_0^2-x\right) (\omega^{(1)}(x)-1)}$ where $x= r_0^2 \sin^2\theta$. Likewise, since $I^{(1)}(\Psi_{\phi}^{(0)})$ the equation for the current can be regarded as
\begin{eqnarray} \label{eq:current}
r_0^2\,I^{(1)}(\Psi_{\phi}^{(0)}) = {2 \,\Psi_{\phi}^{(0)} \left(r_0^2-\Psi_{\phi}^{(0)}\right) (1-\omega^{(1)}(\Psi_{\phi}^{(0)}))}
\end{eqnarray}
The solution for the current $I$ is only one step in find the full solution for the 5D model. To specify the collimated jet we still need to determine the behavior of angular velocity of the magnetic field $\omega_{\phi}$. As an alternative to solving the FFE equations, fortunately, we find that the convergence condition can be applied to get the further details about this function.

Before solving the FFE equations one can analyze the fluxes defined in (\ref{power}) and (\ref{angpower}).
\begin{eqnarray}\label{eq:current}
{P}, {L}_{\phi} \propto I
\end{eqnarray}
On the surface $r_0^2-\Psi_{\phi}^{(0)}=0$ both energy and angular momentum fluxes vanish. On this boundary, at $r\sin\theta=r_0$, the fluxes cannot penetrate. The boundary of the jet is the last field line within the cylinder defined by this boundary. The energy-momentum equations can be divided into two constraint equations.

 In this way we aim to construct a highly-collimated and magnetically-dominated jet solution in the vicinity of spinning 5D (singly spinning) black hole. An interior region $r\sin\theta<r_0$ with a current and field angular velocity
\begin{eqnarray}\label{eq:current}
I \ne 0 \,\,,\qquad \text{and} \qquad \Omega \ne 0  \qquad  (r \sin\theta<r_0)
\end{eqnarray}
 and exterior region where we choose
\begin{eqnarray}\label{eq:current}
I =0 \,\,,\qquad \text{and} \qquad \Omega=0  \qquad  (r \sin\theta>r_0)
\end{eqnarray}
The global solution across the interface at $r\sin\theta=r_0$ will be required to be continuous. In the following we will see that our choice naturally imposes this smoothness condition.
 We to order can now dive into the FFE equations, which to  $O(\alpha^2)$ order can be reduced
\begin{eqnarray}\label{eq:current}
\mathcal{{L}}_{\phi}\Psi_{\phi}^{(2)}=\begin{cases}
    S_{out}, & \text{if $r \,\sin \theta > r_0$}.\\
    S_{in}, & \text{otherwise}.
  \end{cases}
\end{eqnarray}
Here the source term in the outer region ($r \sin \theta > r_0$) yields
\begin{eqnarray}\label{eq:current1}
S_{out}=\frac{4  \, r_0^2  \sin ^2\theta \,(r^4+(r_0^4-3 r^2 r_0^2)\cos^2\theta)}{r^2(r^2-r_0^2)}\,,
\end{eqnarray}
and in the inner region ($r \sin \theta < r_0$) is
 \begin{eqnarray}\label{eq:current2}
S_{in}&=& \frac{r^2  ({I^{(1)}}) ({I^{(1)}})' }{\, (r^2-r_0^2)\, \cos^2\theta}+\frac{4 r^2  \sin ^4\theta \left( r^2-r_0^2 \sin^2 \theta \right)  \left( r_0^4-r^4 {\omega^{(1)}} \right) ({\omega^{(1)}})' }{r_0^2(r^2-r_0^2)}\\
&&- \frac{4 \,  \,r^4  \sin ^2\theta \left( r^2- 2\,r_0^2 \sin^2 \theta \right)( {\omega^{(1)}})^2}{r_0^2(r^2-r_0^2)}+\frac{8 \, \,r^2 r_0^2  \sin ^2\theta \,\cos2\theta \,{\omega^{(1)}}}{(r^2-r_0^2)}\\
&&+\frac{4 \, \, r_0^2  \sin ^2\theta \,(r^4+(r_0^4-3 r^2 r_0^2)\cos^2\theta)}{r^2(r^2-r_0^2)}\,.
\end{eqnarray}
The prime denotes differentiation with respect to $\Psi_{\phi}^{(0)}$. Discussion of the solution of equation (\ref{eq:current}) now follows. If we assume that the convergence condition is true, then we can specify a second relation between $\omega^{(1)}$ and $I^{(1)}$, thus determining the outgoing energy flux. 

As in \cite{Blandford:1977ds}, given a Green's function solution of the differential equation $\mathcal{{L}}_{\phi}G=\delta(r-r_i)\delta(\theta-\theta_i)$, this tells us that we can write down the solution to (\ref{eq:current})
\begin{eqnarray}
\Psi_{\phi}^{(1)}=\int \, r_i \int \, d\theta_i \, G(r,\theta,r_i,\theta_i) S(r_i,\theta_i)
\end{eqnarray}
if, and only if
\begin{eqnarray}
\int_{r_0}^{\infty}dr\int_\delta^{\pi-\delta} d\theta  \,\frac{S_{in/out}}{r}
\end{eqnarray}
converges, where we have defined $\delta=\arcsin(r_0/r)$. The contribution from all terms in (\ref{eq:current1}) and (\ref{eq:current2}) are convergent, assuming ${\omega^{(1)}}\sim O(1)$, except the ones listed here
\begin{eqnarray}
 \frac{r^2 \, r_0^2 ({I^{(1)}}) ({I^{(1)}})' }{\, \cos^2\theta}-{4  \,r^8\,  \sin ^4\theta\,  {\omega^{(1)}} ({\omega^{(1)}})' }- {4 \,r^6  \sin ^2\theta  \,({\omega^{(1)}})^2}=0
\end{eqnarray}
which are now required to vanish for convergence. This equation can alternatively be written as
\begin{eqnarray}
\left(({I^{(1)}})^2- 4 (\omega^{(1)})^2  (\Psi_{\phi}^{(0)})^2\right)'=0\,,
\end{eqnarray}
and integrated to satify
\begin{eqnarray}
({I^{(1)}})^2- 4 (\omega^{(1)})^2  (\Psi_{\phi}^{(0)})^2=\text{constant}\,,
\end{eqnarray}
The integration constant vanishes if we further impose $\Psi_{\phi}^{(0)}=0$ at the polar axis. Combining with the equation for the current (\ref{eq:current}) one can solve explicitly the angular field velocity 
\begin{eqnarray}
\omega^{(1)}=\frac{r_0^2-x}{2 \, r_0^2-x}\,.
\end{eqnarray}
where $x= r^2 \sin^2\theta$. This second relation effectively fixes the current flowing through the hole. Alternative electromagnetic conditions at infinity would yield an alternative relation.
 We have obtained all quantities that are of physical interest about the energy flux. We may  explicitly write down the solution for $\Psi_{\phi}^{(2)}$ but we do not plan to do so because this would only provide us with information about the distortion of magnetic the details about
 we may explicitly write 
 
\begin{figure*}
\centering
	\includegraphics[width=8cm]{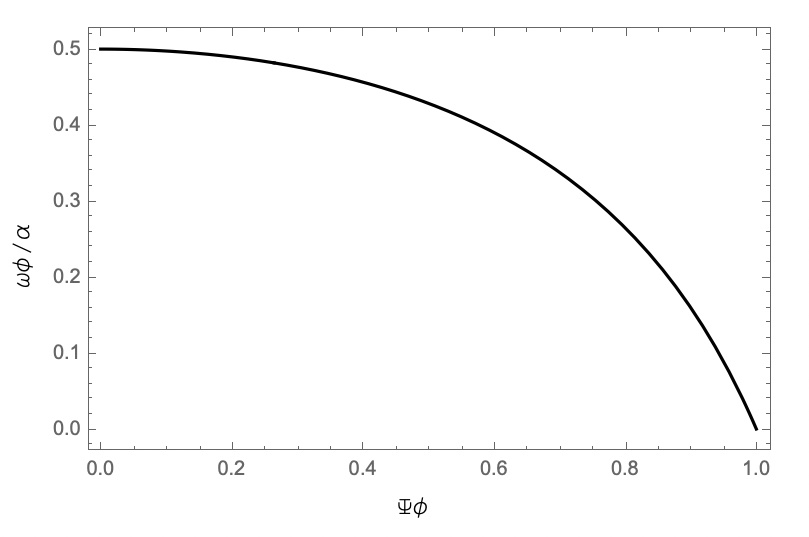}
	\includegraphics[width=8cm]{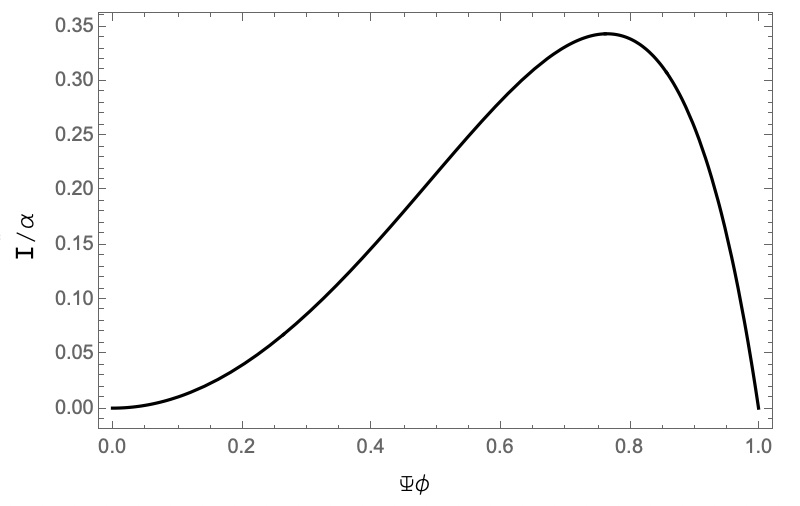}
\caption{Variation of the angular velocity $\omega_{\phi}$ and current $I$ on the black hole horizon.}
\label{fig2}
\end{figure*}

 In Fig. \ref{fig2}, we show the variation the angular field velocity $\omega_{\phi}(\Psi_{\phi})$ and total electric current $I(\Psi_{\phi})$
on the horizon. The behavior of these functions resembles that of 4D spacetimes.
 
 \subsection{Physical Properties of Collimated Jets in 5D}
 \label{subsec:physicalProps}
 
 We have obtained the explicit analytical expressions of the angular velocity $\omega_{\phi}$ and current $I$ for a collimated jet solution in 5D black holes. This leads for the BZ model in 5D black holes to the basic prediction that the power scales as the spin squared as we now explicitly show.
 
\subsubsection*{Energy Flux}

The total energy flux is defined as (\ref{power}) and the angular momentum fluxes as (\ref{angpower}). Equivalently a direct integration leads to
\begin{eqnarray}
P&=& -2 \,(2\pi)^2 \int_0^{r_0^2} I \, \omega_{\phi} \, d\Psi_{\psi} \sim 0.137 \, (2\pi)^2 \,  r_0^4 \, \alpha^2 \,,\\
L_{\phi}&=& -2\, (2\pi)^2 \int_0^{r_0^2} I \, d\Psi_{\psi} \sim 0.455 \, (2\pi)^2 \,  r_0^5 \, \alpha\,.
\end{eqnarray}
The energy extraction efficiency can be employed for the comparison between analytic solutions for the collimated jet in 5D black holes that we derived and the 4D Kerr black hole. 
To keep the amount of magnetic flux crossing the horizon identical we defined the energy extraction efficiency as in \cite{Blandford:1977ds} via
\begin{eqnarray}
\bar{\epsilon}=\frac{\int I \,\omega_{\phi} \,d\Psi_{\psi}}{\int I \,\Omega^H_{\phi}\, d\Psi_{\psi}}
\end{eqnarray}

In 4D the BZ models for collimated jets \cite{Pan:2014bja} display an energy extraction efficiency 
\begin{eqnarray}
\bar{\epsilon}\sim 0.36 \text{ in 4D black holes}
\end{eqnarray}
For the collimated jet solution in Section \ref{Sec:SlowRotation} we obtain
\begin{eqnarray}
\bar{\epsilon}\sim 0.30 \text{ in 5D black holes}
\end{eqnarray}
A direct comparison shows that the collimated jet power is reduced by a factor about $17 \%$ compared to the 4D collimated jet BZ model.

\subsubsection*{Light surface}

\begin{figure}[h]
 \begin{center}
\includegraphics[width=8cm]{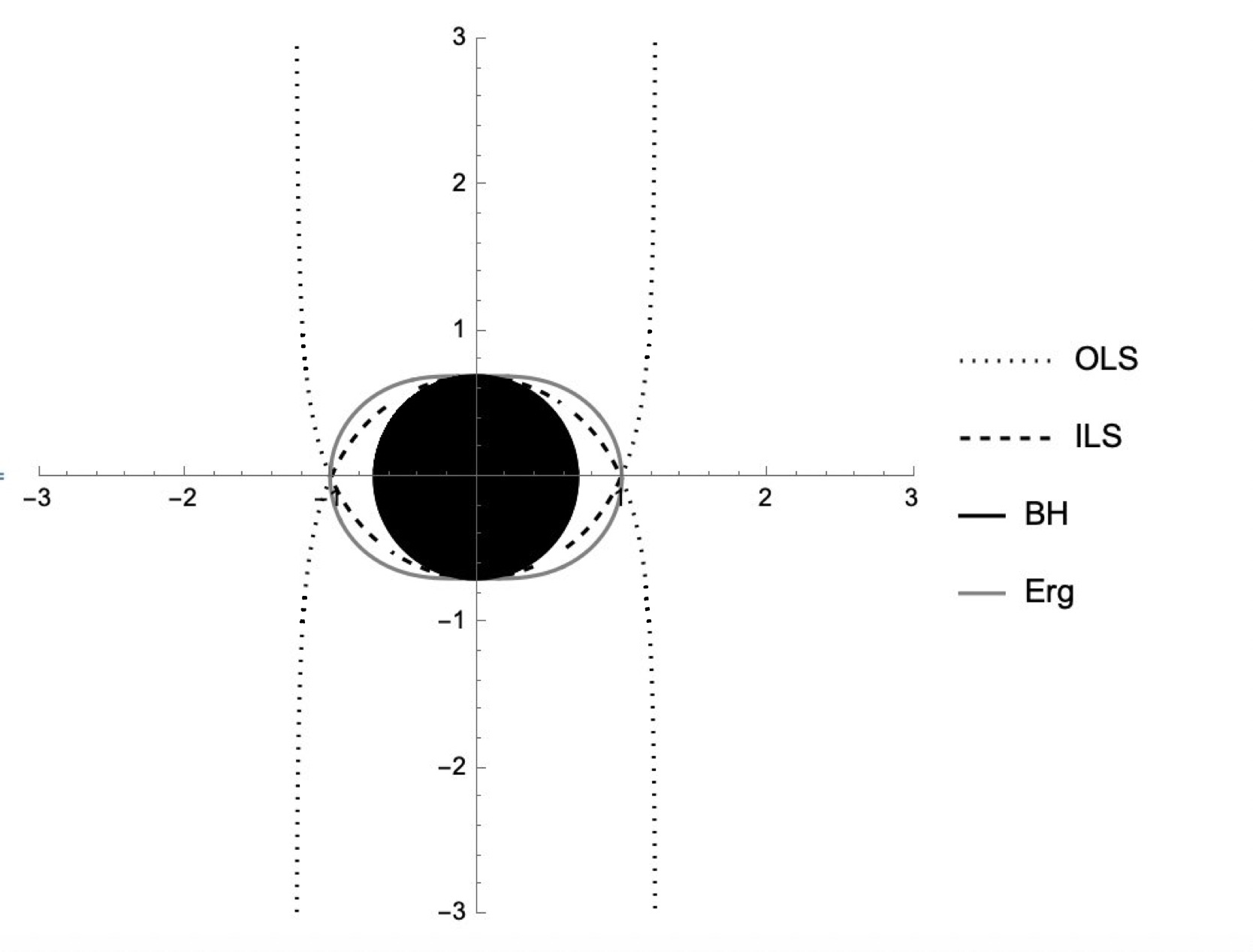}
\caption{Regions in MP black hole: the central {\it black shaded} disk represent the single spinning MP black hole (BH), the {\it dotted} lines correspond to the outer (OLS), the {\it dashed} curve represents the inner light surfaces (ILS), and the ergosphere (Erg) {\it solid} curve for the mass $m=1$ and the spin parameter $\alpha=0.7$.}
\label{fig3}
\end{center}
\end{figure}

Without knowing the precise functional expression of $\omega_{\phi},\omega_{\psi}$ the position of the light surfaces cannot be determined a priori. In the previous section we found the leading order contribution to the angular field velocity, so we will take this solution to depict the positions of the light surfaces for the collimated jet in 5D black holes. We summarize all these finding of the different regions in the space-time in Fig. \ref{fig3}.

There are four critical surfaces characterizing force-free magnetospheres around Myers-Perry black holes: the event horizon, the ergosphere and two light surfaces -- see Section \ref{sec:eventandergo} for details on the first two surfaces. The spacetime location of the light surfaces can be determined by looking at those surfaces where the velocity vector field $\chi= \partial_t+\omega_{\phi} \,\partial_\phi+\omega_{\psi}\, \partial_\psi$ of an observer co-rotating with the magnetosphere becomes null:
\begin{eqnarray}
\chi^2= g_{tt}+ 2\, \omega_{\phi} \,g_{t\phi}+ 2\, \omega_{\psi} \, g_{t\psi}+  \omega_{\phi}^2 \,g_{\phi\phi}+  \omega_{\psi}^2 \,g_{\psi\psi}=0\,.
\end{eqnarray}
Note that the vertical field configuration of our model only crosses the inner light surface. This is a non-trivial interesting case in which there is only one light surface is the uniform vertical configuration, where outside the cylindrical separatrix the field lines are assumed not to rotate. Other more general configurations are expected to cross both light surface (e.g. the radial monopole magnetic field). Since each of these LS is a singular surface of the FFE equations, one can then impose corresponding regularity conditions on these two surfaces, and employing the matched asymptotic expansion, build new FFE solutions for rotating black holes.

\section{Discussion}
\label{Sec:Discussion}

We have shown that one can construct 5D black hole magnetospheres within the FFE approach. To this end we gave analytic solutions of the force-free equations for several model cases. These include explicit 5D solutions in flat spacetime, static black holes and slowly rotating black holes.

As an example, we explicitly construct a highly-collimated and magnetically-dominated jet solution in the vicinity of slowly (single) spinning 5D black holes. 
We propose a general method for perturbative solutions to five dimensional BZ mechanism with a vertical electromagnetic field configuration. Instead of solving the nonlinear FFE equations directly, we rely on the horizon regularity boundary condition, and the convergence requirement to determine relevant physical quantities. This jet solution distinguishes itself from prior known analytic solutions in that it is highly collimated and asymptotically approaches a magnetic cylinder in higher spacetime dimensions. Our solutions confirm and generalize the pioneering results of BZ to higher dimensional spacetimes.

We gave a general argument and then illustrated the power extraction from rotating 5D black hole magnetospheres. The power of energy extraction from the 5D black hole, compared to the predictions of four-dimensional FFE, holds the same scaling with the spin parameter. One sees directly that for the vertical field configurations in higher dimensions does not introduce in the jet power an even steeper scaling on the spin parameter. Whether steeper behaviors in other field configurations are possible (such as in the monopole, parabolic, hyperbolic configurations) remains an open question.

Key to enabling energy extracting magnetospheres was the existence of exact electromagnetic field solutions in static  black holes that we unveil. In fact, we find an infinite tower of FFE configurations in static black hole backgrounds. A possible strategy to build energy extracting magnetospheres for these new field configurations in 5D includes implementing a matched asymptotic expansion (MAE) \cite{Camilloni:2022kmx,Armas:2020mio,Gralla:2015vta}. Preliminary attempts in this direction have shown that one has to consider magnetic field lines that extend from the event horizon out to infinity and now have to cross two light surfaces (the inner and the outer light surfaces) in five spacetime dimensions. Thus, we propose that one should be able to devise an iterative scheme that uses regularity on the light surfaces to determine the two free functions -- the current $I$ and angular field velocity $\omega_{\phi}$. It is worth emphasizing that there is a problem for a vertical field configuration within the MAE approach. All magnetic field lines cross the inner light surface, but none crosses both light surfaces. Therefore, we cannot implement the MAE method in which one determines solutions to the FFE equations through the condition of smooth crossing of both light-surfaces. Thus here we resort to set regularity on the horizon and the location of zero power surfaces.  

The interaction of black holes with ambient magnetic fields is important for a variety of highly energetic astrophysical phenomena. Previous studies of the FFE black hole magnetospheres mechanism outside traditional BZ model in GR include \cite{Callebaut:2020hwv,Dong:2021yss,Iqbal:2021ouw}. Our goal in 5D spacetimes is a deeper understanding of the structure of the rotating black hole magnetosphere.
 
We emphasize that all the above examples are certainly simplified. Nevertheless, the solutions obtained are capable of creating several key features that allow one to judge the basic properties of the central engine. In Kerr black holes the so called ‘Meissner effect’ could potentially quench jet power at the highest black hole spins values \cite{Komissarov:2007rc,Penna:2014aza}.  We believe that in a force-free plasma filled environment in higher dimensions no Meissner effect will ever occur for any black hole with one vanishing spin. On the one hand, the 5D the extremal black hole solution is singular and the BZ model at the highest spin value $m=a^2$ will be plagued with singularities. On the other hand, in $D>5$ black objects can overspin and hence never become extremal. An interesting by product is that magnetospheres in $D>5$ over-spinning black holes may possibly tap infinite energy from the system.

As an extension of our work we would like to investigate FFE solutions in 5D black hole backgrounds with two angular momenta, in contrast to one non-trivial spin black hole background presented here. We expect a rich field structure for the FFE models in these generic Myers-Perry black holes. It is yet to be determined whether the problem contains FFE solutions that lead to power extraction.

GR in 5D contains also non-spherical black hole solutions, namely black rings with event horizon topology $S^1\times S^2$ \cite{Emparan:2006mm,Pomeransky:2006bd}. This is not the only example of non-spherical horizon topology, that also turned out to be a counterexample to black hole uniqueness. Examples include bicycling black rings \cite{Elvang:2007hs} and other more elaborate black hole configurations -- see \cite{Rodriguez:2010zw} for a review -- are also known analytically in closed form.  Therefore, another interesting scenario is the extension of the BZ models in black ring-like backgrounds. We argue that configurations of this sort will not support monopole fields, but instead magnetic dipole electromagnetic fields.

Finally, let us comment on the FFE in the zoom in regions in 5D black holes. The near horizon extremal black hole geometries display symmetry enhancement, and so does its dynamics \cite{Lupsasca:2014pfa,Zhang:2014pla,Camilloni:2020qah}. We therefore expect that 5D BZ model in the near horizon extremal black holes geometries to emulate the lower dimensional settings, and carry infinite towers of FFE solutions.

To summarize, we showed herein, for the first time developed a consistent FFE model for black hole jets in 5D, and were able to establish some of the effects of the number of spacetime dimensions. In particular, we revealed that a black hole in five space-time dimensions can support energy extracting magnetospheres in form of vertical collimated jets. These 5D models are not expected to be realize in nature, however, these represent new insights of rotating black hole magnetosphere, the strong-gravity regime in GR and signatures of extra dimension in black hole jets.

\section*{Acknowledgement}
The work of ABC is partially supported through a Utah Nasa Space Consortium Grant fellowship, as well a USU Howard L. Blood Fellowship. The work of MJR is partially supported through the NSF grant PHY-2012036, RYC-2016-21159,  he grants CEX2020-001007-S and PGC2018-095976-B-C21, funded by MCIN/AEI/10.13039/501100011033.  
\appendix

\section{FFE solutions for $5D$ flat space-times}
 \label{app:}
When $m=0$ and $a=0$ the solution $(\ref{mphole})$ corresponds to $5D$ flat solution. In this section we find solutions to FFE equations in flat space-time.

\subsubsection*{Radial Solution}
The following field strength satisfies the force-free equations in flat 5D space-time
\begin{eqnarray}
F= \,d\theta \wedge\left[ c_1 \tan\theta \, d\phi-\frac{1}{\sin\theta\cos\theta}\,\left(\frac{c_3 }{r}dr\pm  c_2 \, dt \right)\right]\,.
\end{eqnarray}
for arbitrary values of the constants $c_i$ with $i=1,2,3$. The solution has $dF=0$ and also is degenerate $F\wedge F=0$.

\subsubsection*{Non-closed Radial Solution}
As a first and immediate generalization of the known exact solutions to force-free equation in 4D, is one that has radial fluxes. We make the same assumptions for the angular velocity of the field and current $I$. A solution of the form $(\ref{eq:fieldstrength})$ that has radial fluxes $\Psi_{\phi,\psi}$ independent of $r$, is
\begin{eqnarray}
\Psi_{\phi}=1-c_1 \log(\cos\theta)\,,\qquad \Psi_{\psi}=1-c_2 \log(\sin\theta)\,,\\
\omega_{\phi}=\frac{\pm \,I(\theta)}{2(\partial_{\theta}{\Psi_{\phi}})}\sqrt{ \frac{\det g_2}{-\det g_1}} \,,\qquad \omega_{\psi}=\frac{\pm \,I(\theta)}{2(\partial_{\theta}{\Psi_{\psi}})}\sqrt{ \frac{\det g_2}{-\det g_1}}\,.
\end{eqnarray}
for an arbitrary function of $I(\theta)$ and $c_1,c_2$ constants where $\sqrt{ \frac{\det g_2}{-\det g_1}}= (r\sin\theta\cos\theta)^{-1}$. This solution is however discontinuous at the pole $\theta=0$ thereby $c_2=0$ (and $c_1=1$) which leads to
\begin{eqnarray}
F= d\theta \wedge\left[ \tan\theta \, d\phi-\frac{I(\theta)}{r\sin\theta\cos\theta}\,(dr\pm dt)\right]\,.
\end{eqnarray}
In this way, the effective solution that we will consider, has fluxes (discontinuous at the equator $\theta=\pi/2$) and electromagnetic angular momenta only in the $\phi$-direction and takes the form
\begin{eqnarray}
\Psi_{\phi}=1-c_1 \log(\cos\theta)\,,\qquad \Psi_{\psi}=0\,,\label{eq:BCinfity1}\\
\omega_{\phi}=\frac{\pm \, I(\theta)}{r\sin\theta\cos\theta\, (\partial_{\theta}{\Psi_{\phi}})}\,,\qquad \omega_{\psi}=0\label{eq:BCinfity2}\,.
\end{eqnarray}
for an arbitrary function of $I(\theta)$ with $c_1$ constants. This solution is a generalization to $5D$ of Michel's solution which also is a radial solution. Moreover, the relation for $\omega_{\phi}$ in (\ref{eq:BCinfity2}) is the extension of the four-dimensional boundary condition for the fields $\omega^{4D}=\pm \, I^{4D}\sqrt{ \frac{\det g^{4D}_2}{-\det g^{4D}_1}}\,(\partial_{\theta}{\Psi^{4D}})$ where $\sqrt{ \frac{\det g^{4D}_2}{-\det g^{4D}_1}}= (\sin\theta)^{-1}$. See e.g. equation (2.18) in \cite{Gralla:2015vta}. The $\pm$ signs in the solution represent ingoing/outgoing flux.

Under similar assumptions to 4D, the force-free equations can be satisfied in 5D flat space-time. Notice however that in 5D the field strength solution does not close $dF\ne0$ (but is degenerate  $F\wedge F=0$) for an arbitrary current function $I(\theta)$. We therefore consider the solution not physical, and regard it as a reminder that energy extraction process in 5D will include new features when comparing to the 4D BZ model.


\end{document}